\documentclass[12pt,tightenlines,eqsecnum,floats,aps,amsmath,amssymb,nofootinbib,prd]{revtex4}

\usepackage{setspace}
\usepackage{subfig}
\usepackage{amsmath,amssymb,amsfonts,amsthm,mathrsfs}
\usepackage{graphicx}
\usepackage{enumerate}

\def\e{{\bf e}}                                                       
\def\a{{\bf a}} 
\def\k{{\bf k}} 
\def\F{{\bf F}} 
\def\C{{\bf C}} 
\def\g{{\pmb \gamma}}

\begin{document}

\title{Classical axisymmetric gravity in real Ashtekar variables}

\author{Rodolfo Gambini$^{1}$, Esteban Mato$^{1}$, Javier
  Olmedo$^{2,3}$, Jorge Pullin$^{3}$}
\affiliation {
1. Instituto de F\'{\i}sica, Facultad de Ciencias, 
Igu\'a 4225, esq. Mataojo, 11400 Montevideo, Uruguay. \\
2. Institute for Gravitation and the Cosmos, Penn State University,  
University Park, PA 16801, USA\\
3. Department of Physics and Astronomy, Louisiana State University,
Baton Rouge, LA 70803-4001, USA}

\begin{abstract}
We formulate axisymmetric general relativity in terms of real
Ashtekar--Barbero 
variables. We study the constraints and equations of motion and show
how the Kerr, Schwarzschild and Minkowski solutions arise. We also
discuss boundary conditions. This opens the possibility of a
midisuperspace quantization using loop quantum gravity techniques for
spacetimes with axial symmetry and time dependence. 
\end{abstract}
\maketitle
\section{Introduction}

Due to the complexities of the quantized version of the Einstein
equations in loop quantum gravity, the study of mini and
midisuperspaces has proved a valuable tool to gain insights into the
physics of the theory. The study first started with homogeneous
cosmologies, giving rise to loop quantum cosmology (see
\cite{Ashtekar:2011ni} and references therein). It was later expanded
to include spherically symmetric spacetimes (see
\cite{Gambini:2013hna} and references therein), including charged
black holes \cite{Gambini:2014qta}. In both cases interesting physical
insights, like the elimination of singularities due to quantum
effects, were found. It is natural to try to extend these studies to
situations with less symmetry, like the case of axisymmetric
spacetimes, which include physically important situations, like the
Kerr geometry. There is virtually no literature on the subject. An
exception is the work on isolated horizons and black hole entropy
\cite{krasnov:1998,bojowald:2000,perez:2011,bianchi:2011,frodden:2014,achour:2016,croken:2017}. An early
study of spacetimes with one Killing vector field made some progress,
partially addressing the situation of axial symmetry in complex connection variables
\cite{Husain:1989mp}. Some progress was also made in planar
space-times (\cite{Neville:2013xba} and references therein,
\cite{Hinterleitner:2011rb}) and the case of two spatial Killing
vector fields was also discussed for the Gowdy models
\cite{Husain:1989qq}, including the use of hybrid quantizations (see
\cite{ElizagaNavascues:2016vqw} for a review). Some of these studies
were in terms of the early form of the Ashtekar variables which were
complex.

Here we would like to discuss the case of axisymmetric space-time
using real Ashtekar--Barbero variables. We introduce a suitable
Killing
vector field and coordinates adapted to it.
We will also show how the Kerr,
Schwarzschild and Minkowski solutions arise. Besides, we will make some
remarks on boundary conditions. This completes a classical setup
suitable to perform a loop quantization, which we will discuss in a
subsequent paper. This is the first example of a system with only one
Killing vector field to be formulated with the real Ashtekar--Barbero
variables.

The organization of this paper is as follows. In section 2 we discuss
a set of symmetry adapted variables and set up the kinematics of the
problem. In section 3 we introduce the constraints of general
relativity in terms of the reduced axisymmetric variables introduced. In
section 4 we work out the equations of motion. In section 5 we check that some particular
solutions of interest including the Kerr, Schwarzschild and Minkowski
space-times solve the equations we present. In section 6 we discuss
boundary conditions. We end with a discussion.

\section{Kinematics: symmetry adapted variables}

Here we will impose a symmetry reduction due to a spatial Killing
field with orbits tangent to $S^1$. Let us consider a choice of
fiducial coordinates $\{x,y,\phi\}$, where $\phi\in S^1$ and
$x,y\in \mathbb{R}$. The Killing field will be then
\begin{equation}
K^a=\left(\partial_\phi\right)^a.
\end{equation}

We will be following the typical reduction procedure adopted for
connection variables. Namely, a connection
$ A=A_a^i\tau_i \mathrm{d}x^a$ will be invariant under the Killing
symmetries if it satisfies the condition
\begin{equation}
{\cal L}_{\tilde K}A^i_a=\epsilon_{ijk}\lambda^jA^k_a,
\end{equation} 
where $\tilde K = \lambda_i \partial^i=\lambda_3\partial_\phi$ and $\lambda_1=0=\lambda_2$. The previous equation amounts to
\begin{equation}\label{eq:axi-sym}
\partial_\phi A^i_a=\epsilon_{i3k}A^k_a.
\end{equation} 
Notice that we are imposing that the Lie derivative be proportional to
a constant $O(2)$ gauge transformation \cite{cordero}. We have found
this to be the simplest choice that is general enough to recover all
solutions with axisymmetry. In other situations one may need to
consider $\lambda^i$ that are more general, perhaps including spatial
dependence. 

The same equation is valid for the densitized triad $E^a_i$. The most general solution (see Appendix \ref{app:red}) to these equations are
\begin{eqnarray}\label{eq:red-A}
 A&=&A_a^i\tau_i \mathrm{d}x^a= \left((\cos(\phi)\tau_1 + \sin(\phi) \tau_2)  \a_a^1 + (-\sin(\phi) \tau_1 + \cos(\phi) \tau_2) \a_a^2 + \a_a^3 \tau_3 \right)\mathrm{d}x^a \\ \label{eq:red-E}
 E&=&E^a_i \tau^i \partial_a= \left((\cos(\phi)\tau^1 + \sin(\phi) \tau^2) \e^a_1 + (-\sin(\phi) \tau^1 + \cos(\phi) \tau^2) \e^a_2 + \e^a_3 \tau^3 \right)\partial_a,
\end{eqnarray}
where the symmetry adapted variables $(\a_a^i,\e^b_j)$ do not depend
on the angular coordinate $\phi$, i.e. only on $(x,y)$, and are
canonically conjugate. In order to prove this, it is very easy to verify
that
\begin{equation}
  \Omega=\frac{1}{8\pi G \beta}\int dx dy d\phi \;\delta E^a_i\wedge \delta A^i_a = \frac{1}{4 G \beta}\int dx dy \;\delta \e^a_i\wedge \delta\a^i_a,
\end{equation}
with $\beta$ the Immirzi parameter. In other words,
\begin{equation}
\{\a^i_a(\vec x),\e^b_j(\vec x')\}=4 G\beta\,\delta^i_j\delta^b_a\delta^{(2)}(\vec x-\vec x'). 
\end{equation}
Another geometrical quantity that will be useful and can be computed now is the determinant of the symmetry-reduced densitized triad
\begin{equation}
E = \det(E) =\frac{1}{3!}\varepsilon_{abc}\varepsilon^{ijk}E^a_iE^b_jE^c_k=\frac{1}{3!}\varepsilon_{abc}\varepsilon^{ijk}\e^a_i\e^b_j\e^c_k=\det (\e)=\e. 
\end{equation}
The inverse of the densitized triad, $E^i_a$, takes a similar form as
$E^a_i$, but replacing $\e^a_i$ by $\e_a^i$. One can easily see that
$\e^i_a$ fulfills $\e^i_a\e^a_j=\delta^i_j$ and
$\e^i_a\e^b_i=\delta^a_b$, i.e. it is the inverse of $\e^a_i$ (and
therefore it can be written in terms of $\e^a_i$). Then, the  symmetry-reduced spatial metric can be written as
\begin{equation}
q_{ab}=EE^i_aE^i_b=\e\,\e^i_a\e^i_b,
\end{equation}
and it only depends on $\e^a_i$. 

Similarly, the same reduction process can be applied to the extrinsic curvature $ K=K_a^i\tau_i \mathrm{d}x^a$, in its triadic form, and the spin connection $ \Gamma=\Gamma_a^i\tau_i \mathrm{d}x^a$, namely
\begin{eqnarray}
 K&&=\left((\cos(\phi)\tau_1 + \sin(\phi) \tau_2)  \k_a^1 + (-\sin(\phi) \tau_1 + \cos(\phi) \tau_2) \k_a^2 + \k_a^3 \tau_3 \right)\mathrm{d}x^a,\\
 \Gamma&&=\left((\cos(\phi)\tau_1 + \sin(\phi) \tau_2)  {\gamma}_a^1 + (-\sin(\phi) \tau_1 + \cos(\phi) \tau_2) \gamma_a^2 + \gamma_a^3 \tau_3 \right)\mathrm{d}x^a.
\end{eqnarray}
Actually, the components of the symmetry-reduced spin connection can be written in terms of the components of the symmetry-reduced triads as
\begin{equation}                                                                
\gamma^i_a=\frac{1}{2}\epsilon_{ijk}\e^b_j\left(\e_{a,b}^k-\e_{b,a}^k+\e^c_k\e_a^l\e^l_{c,b}+\e_a^k\e^l_{c}\e_{l,b}^{c}\right)-\delta^i_3\delta^\phi_a=  \g^i_a-\delta^i_3\delta^\phi_a,                                       
\end{equation}
where $\g^i_a$ is the spin connection compatible with $\e^a_i$. This means that $\{\e^a_i,\g^j_b\}=0$, and therefore $\{\e^a_i,\gamma^j_b\}=0$.

The relation between the components of the symmetry-reduced extrinsic curvature with the ones of the symmetry-reduced Ashtekar-Barbero and the spin connections is
\begin{equation}
\beta\k^i_a=\a^i_a-\gamma^i_a = \a^i_a-\g^i_a+\delta^i_3\delta^\phi_a. 
\end{equation}

Finally, we will conclude this section by introducing some identities
that can be useful for the calculations in the next sections. The
first identity is the Poisson bracket of the connection with the
inverse triad,
\begin{equation}
\{\a^i_a(\vec x),\e^j_b(\vec x')\}=-4 G\beta\delta^{(2)}(\vec x-\vec x')\, \e^i_b\e^j_a.
\end{equation}
and it is easy to prove. The second identity, 
\begin{equation}
\{\a^i_a(\vec x),\e(\vec x')\}=4 G\beta\delta^{(2)}(\vec x-\vec x')\, \e\,\e^i_a,
\end{equation}
is based on $\delta\e = \e \,\e^i_b \delta \e_i^b$, for any given
variation $\delta\e$ (in particular it is therefore valid for
$\partial_a\e$). For the next identity, we should first notice that,
given any phase space tensor of the form $T^a_i$ with density weight
one, we can define
\begin{equation}
\g(T)=\int dx dy \,T_i^a\g^i_a,
\end{equation}
at least formally (we are not taking into account boundary terms
neither fall-off conditions). These types of expressions appear, for
instance, in the Lorentzian part of the symmetry-reduced Hamiltonian
constraint. After a lengthy but simple calculation, one can prove that
\begin{equation}\label{eq:PB-ident}
\{\a^i_a(\vec x),\g(T)\}=4 G \beta \left.{P^{ij}}{}^b_{ac}{}^{\e}\!D_bT^c_{j}\right|_{\vec x}+\int dx' dy' \,\{\a^i_a(\vec x),T_j^b(\vec x')\}\g^j_b(\vec x'),
\end{equation}
where 
\begin{equation}\label{eq:PB-P-def}
{P^{ij}}{}^b_{ac}=\frac{1}{2}\left[\epsilon_{i j k}\e^{l}_c\left(\e^{l}_a   \e_{k}^b +  \e_{l}^b \e^{k}_a\right) + \epsilon_{ljk} \e_{k}^b\left(\e^i_c   \e^{l}_a  - \e^i_a \e^{l}_c \right)\right]   
\end{equation}
and ${}^{\e}\!D_b$ is the covariant derivative compatible with $\e^a_i$, namely ${}^{\e}\!D_b\e^a_i=0$. 

Similar identities also hold in the full theory, namely, for the original phase space variables $A^i_a$ and $E_j^b$, and the spin connection $\Gamma^i_a$.

\section{The constraints}

The total Hamiltonian of the full theory is a combination of 7
constraints: 3 Gauss constraints, 3 vector constraints and
the Hamiltonian constraint. Concretely,
\begin{equation}
H_T=\frac{1}{16\pi G}\left[G(\vec\Lambda)+ D(\vec N)+C(N)\right],
\end{equation}
where
\begin{align}
G(\vec\Lambda)&=\frac{2}{\beta}\int d^3x \Lambda^i\left(\partial_aE^a_i+\varepsilon_{ijk}A^j_aE^a_k\right),\\
D(\vec N)&=\frac{2}{\beta}\int d^3x N^a\left(E^b_i\partial_aA^i_b-\partial_b(E^b_iA^i_a)\right), \\
C(N) &= H_E(N)+H_L(N),
\end{align}
and where $H_E(N)$ and $H_L(N)$ are the Euclidean and Lorentzian parts of the Hamiltonian constraint, given, respectively, by
\begin{align}
H_E(N)&=-\int d^3x N e^{-1}(A^{i}_{b,a} - A^{i}_{a,b} + \epsilon_{ilm}A^{l}_{a} A^{m}_{b})\epsilon_{ijk}E_{j}^{a} E_{k}^{b} ,\\
H_L(N)&=\int d^3x N (1+\beta^2)e^{-1}\epsilon_{ijk} \epsilon_{ilm} E_{j}^{a} E_{k}^{b} K^{l}_{a} K^{m}_{b}.
\end{align}

Now, we replace the symmetry-reduced connection $A^i_a$ and the densitized triad $E^a_i$ in the previous expressions. The symmetry-reduced Hamiltonian will be
\begin{equation}
h_T=\frac{1}{8 G}\left[g(\vec\Lambda)+ d(\vec N)+c(N)\right],
\end{equation}
and the (smeared) constraints are,
\begin{align}
g(\vec\lambda)&=\frac{2}{\beta}\int dxdy \lambda^i\left(\partial_a\e^a_i+\varepsilon_{ijk}\a^j_a\e^a_k+\varepsilon_{ijk}\delta^j_3\delta^\phi_a \e^a_k\right),\\
d(\vec N)&=\frac{2}{\beta}\int dxdy N^a\left(\e^b_i\partial_a\a^i_b-\partial_b(\e^b_i\a^i_a)+\delta_a^\phi\delta^i_3\varepsilon_{ijk}\a^j_b\e^b_k%-\delta_a^\phi\partial_b(\e^b_i\a^i_c\delta^c_\phi)
\right), \\
h_E(N)&=-\int dxdy \frac{N}{\sqrt{\e}}\left[(\a^{i}_{b,a} - \a^{i}_{a,b} + \epsilon_{ilm}\a^{l}_{a} \a^{m}_{b})\epsilon_{ijk}\e_{j}^{a} \e_{k}^{b} +2\delta^j_3 \delta_b^\phi\left(\a^i_a \e_i^a\e_j^b - \a^i_a\e_i^b \e_j^a\right)\right] ,\\
h_L(N)&=\int dxdy \frac{N}{\sqrt{\e}} (1+\beta^2)\epsilon_{ijk} \epsilon_{ilm} \e_{j}^{a} \e_{k}^{b} \k^{l}_{a} \k^{m}_{b},
\end{align}
where, as before, we have written the scalar constraint as
$c(N)=h_E(N) + h_L(N)$.

\section{Equations of motion}

In this section we will provide the Poisson brackets of the components
of the symmetry-reduced connection and densitized triad with the
constraints. We will perform a local analysis, assuming suitable
boundary terms have been chosen. We will discuss the issue of boundary
terms for the asymptotically flat case in section 5, similar analyses
can be carried out for other asymptotic behaviors.  Let us start with
the Gauss constraint. One can easily see that
\begin{align}
\{\e^a_i(\vec x),g(\vec\lambda)\} &= \left. \left(\epsilon_{ijk}  \lambda^j \e_k^a\right)\right|_{\vec x} \\
\{\a_a^i(\vec x),g(\vec\lambda)\} &= \left. \left(-\lambda^i_{,a}+\epsilon_{ijk}  \lambda^j \a^k_a+\varepsilon_{ijk}\lambda^j\delta^k_3\delta^\phi_a \right)\right|_{\vec x}.
\end{align}
We see that $\e^a_i$ transforms as a tensor and that
$\a^k_a+\delta^k_3\delta^\phi_a$ transforms as a connection.

The vector constraint yields 
\begin{align}
\{\e^a_i(\vec x),d(\vec N)\} &= \left. \left(\left(N^b \e^a_{i}\right)_{,b}-\e_i^b N^a_{,b}
-\varepsilon_{ijk}N^d\delta_d^\phi\delta^k_3\e^a_j 
%- N^d_{,b}\delta_d^\phi\delta^a_\phi\e^b_i
\right)\right|_{\vec x} \\
\{\a_a^i(\vec x),d(\vec N)\} &= \left. \left(\a^i_b N^b_{,a} + N^b \a^i_{a,b}
+\varepsilon_{ijk}N^d\delta_d^\phi\delta^j_3\a^k_a 
%+ N^d_{,a}\delta_d^\phi\delta^c_\phi\a^i_c
\right)\right|_{\vec x}.
\end{align}
The Poisson brackets with the Euclidean and Lorentzian parts of the
Hamiltonian constraint are,
\begin{align}
\{\e^a_i(\vec x),h_E(N)\} &= - \frac{\beta}{2}\left. \left[2\left(\frac{N}{\sqrt{\e}}\varepsilon_{ijk}\e_{j}^{b} \e_{k}^{a}\right)_{,b}-2 \frac{N}{\sqrt{\e}} \varepsilon_{ilm}   \varepsilon_{mjk} \a^{l}_b   \e_{j}^{a}  \e_k^b - 2\frac{N}{\sqrt{\e}}\delta^j_3 \delta_b^\phi\left(\e_i^a\e_j^b - \e_i^b \e_j^a\right)\right]\right|_{\vec x} \\
\{\e^a_i(\vec x),h_L(N)\} &= \frac{\beta}{2}\left. \left(
\frac{2}{\beta} (1+\beta^2)\frac{N}{\sqrt{\e}} \varepsilon_{ilm}   \varepsilon_{jkm} \k^{l}_{b}   \e_{j}^{b}   \e_k^a   
\right)\right|_{\vec x} 
\end{align}

Finally, we will provide the Poisson brackets of the components of the symmetry-reduced connection with the symmetry-reduced Hamiltonian constraint. The Euclidean part is simply given by
\begin{align}\nonumber
\{\a_a^i(\vec x),h_E(N)\} &= -\frac{\beta}{2}\left[-\frac{N}{2}\C_E\e^i_a+2\frac{N}{\sqrt{\e}}\epsilon_{ijk}\F^k_{ab}\e^b_j\right.\\
&\left.\left.+\frac{2N}{\sqrt{\e}}\left(\delta^j_3 \delta_b^\phi\a^i_a \e_j^b+\delta^i_3 \delta_a^\phi\a^j_b \e_j^b-\delta^j_3 \delta_a^\phi\a^i_b \e_j^b-\delta^i_3 \delta_b^\phi\a^j_a \e_j^b
\right)\right]\right|_{\vec x},
\end{align}
where $\F_{ab}^i= \a^{i}_{b,a} - \a^{i}_{a,b} + \epsilon_{ilm}\a^{l}_{a} \a^{m}_{b}$.
Here we have introduced 
\begin{equation}
\C_E=\frac{1}{\sqrt{\e}}\left[\F_{ab}^i\epsilon_{ijk}\e_{j}^{a} \e_{k}^{b} +2\delta^j_3 \delta_b^\phi\left(\a^i_a \e_i^a\e_j^b - \a^i_a\e_i^b \e_j^a\right)\right],
\end{equation}
that is nothing but the Euclidean part of the local Hamiltonian constraint $\C$. On the other hand, for the Lorentzian part we must notice that $\{\a^i_a,\k^j_b\}=-\beta^{-1}\{\a^i_a,\g^j_b\}$ and also remember the identity in \eqref{eq:PB-ident} and the definition \eqref{eq:PB-P-def}. After some manipulations, one gets,
\begin{align}\nonumber
\{\a_a^i(\vec x),h_L(N)\} &= -\frac{\beta}{2} \left\{-\frac{N}{2}\C_L\e^i_a-2(1+\beta^2)\frac{N}{\sqrt{\e}} \varepsilon_{ijk}   \varepsilon_{lmk} \k^{l}_{a}   \k^{m}_{b}   \e_j^b\right.\\
&+\left.\left.\frac{1}{\beta}(1+\beta^2) \left(\varepsilon_{ijk}\e^m_a-\frac{1}{2}\varepsilon_{mjk}\e^i_a\right)\e_j^b \e_k^c\left[{{}^{\e}\!D_b}\left(\frac{N \k^m_{c}}{\sqrt{\e}}\right)-{{}^{\e}\!D_c}\left(\frac{N \k^m_{b}}{\sqrt{\e}}\right)\right]
\right\}\right|_{\vec x}. 
\end{align}
Similarly, 
\begin{equation}
\C_L=\frac{1}{\sqrt{\e}} (1+\beta^2)\epsilon_{ijk} \epsilon_{ilm} \e_{j}^{a} \e_{k}^{b} \k^{l}_{a} \k^{m}_{b},
\end{equation}
is the Lorentzian part of the Hamiltonian
constraint $C$.

\section{The Kerr, Schwarzschild and Minkowski solutions}

We have explicitly checked that the symmetry-reduced model admits the
well-known solution of the full theory given by the Kerr metric.

In the usual spherical coordinates $(r,\theta,\phi)$, the
densitized triad for Kerr in a diagonal gauge, identifying the
internal directions with the coordinates,  takes  the following form:
\begin{eqnarray}\nonumber
 \e^r_3 &&= \sin\theta \sqrt{(r^2+a^2)(r^2+a^2 \cos^2\theta)+a^2 ~ r ~ r_s \sin^2\theta},\\\nonumber
 \e^\theta_1&&=\frac{\sin\theta \sqrt{(r^2+a^2)(r^2+a^2 \cos^2\theta)+a^2 ~ r ~ r_s \sin^2\theta}}{\sqrt{a^2+r^2-r~r_s}},\\\label{eq:triad}
 \e^\phi_2&&=\frac{r^2+a^2\cos^2\theta}{a^2+r^2-r~r_s},
\end{eqnarray}
where $a=J/r_s$, while the rest of its components vanish. Together
with the well-known choices of lapse and shift (e.g. \cite{visser,chandrasekhar})
\begin{eqnarray}
 N&&=\sqrt{\frac{(a^2+r(r-r_s))(a^2+2r^2+a^2 \cos(2\theta))}{2(a^2+r^2)(r^2+a^2\cos^2\theta+2 a^2 ~r ~r_s \sin^2\theta)}},\\
N_\phi&&=-\frac{r r_s a \sin^2\theta}{r^2+a^2\cos^2\theta},\quad N_r=0,\quad N_\theta = 0,
\end{eqnarray}
and one can easily verify that this solution corresponds to the Kerr
metric in Boyer--Lindquist coordinates. Here, $r_s$ is the
Schwarzschild radius and $a$ the angular momentum per unit mass. The
spin connection can be computed out of the densitized triad. Finally,
the connection components can be easily computed provided the
extrinsic curvature in triadic form. Concretely,
\begin{align}
 k_a^i&=\delta^{ij}\frac{\e^{b}_j}{\sqrt{\e}}K_{ab}
\end{align}
where we obtain the extrinsic curvature from
\begin{align}
 K_{ab}&=\frac{1}{2N}\left( -\dot h_{ab}+\nabla_a N_b + \nabla_b N_a \right),
\end{align}
keeping in mind that for our stationary solution $\dot h_{ab}=0$. The
expressions of the components of the connection are rather lengthy, as
well as those of the Langrange multipliers $\lambda^i$ of the Gauss
constraint. We give them in the appendix. To determine the Lagrange
multipliers we insert the above expressions for the triad, connection,
lapse and shift in the equations of motion. 

It is also very easy to check that in the limit $a\to 0$ we recover the Schwarzschild (static) solution. Concretely, the densitized triad reduces to
\begin{eqnarray}\nonumber 
 \e^r_3&&=r^2 \sin\theta, \\\nonumber
 \e^\theta_1&&=\frac{r\sin\theta}{\sqrt{1-\frac{r_s}{r}}}, \\
 \e^\phi_2&&=\frac{r}{\sqrt{1-\frac{r_s}{r}}}.
\end{eqnarray}
the lapse now takes the familiar form
\begin{align}
 N&=\sqrt{1-\frac{r_s}{r}},
\end{align}
while the shift vanishes, namely $N_\phi=0$. Besides, we also have that $K_{ab}=0$. Therefore, the connection is completely determined by the spin connection.

The Lagrange multipliers for the Gauss constraint  take the simple
form (for the Schwarzschild case, for the Kerr case see the appendix):
\begin{align}\nonumber
 \lambda^1&=0 \\\nonumber
 \lambda^2&=0 \label{lambda_schwarzschild} \\
 \lambda^3&=-\beta \frac{r_s}{r^2}
\end{align}

Finally, the Minkowski solution can be recovered from the limit $a\to 0$ and $r_s\to 0$. The densitized triad reduces  to
\begin{eqnarray}
 \e^r_3&&=r^2 \sin\theta, \\
 \e^\theta_1&&=r\sin\theta, \\
 \e^\phi_2&&=r.
\end{eqnarray}
The lapse function $N=1$ becomes the usual one in flat spacetimes. As
in the previous case, the spin connection completely determines the
connection, and any other Lagrange multiplier (shift and $\lambda^i$)
vanish.

\section{Boundary terms}

Up to now the analysis we made has been local. When one is in
asymptotically flat space-times one needs to be mindful about falloff
rates and integrations by parts. In particular, in addition to the
constraints, one has a true Hamiltonian associated to the generators of
the Lorentz group at infinity. In this section we will identify the
boundary contributions needed to make the action differentiable in the
asymptotically flat case for the Ashtekar--Barbero variables with
axial symmetry. We will review individually each set of constraints to
see if boundary terms are needed. We will follow closely
\cite{thiemann_boundary,campiglia_asymptotic}. 

\subsection{Diffeomorphism constraint}

We start this section by writing the portion of the action that
corresponds to the diffeomorphism constraints in ADM-variables:
\begin{align}
  D[\vec{N}] = - 2 \int_{\Sigma} \mathrm{d}^3 x\,N^a\, \nabla_b P_{~a}^b,
\end{align}
$\nabla$ being  the covariant derivative compatible with the metric $g_{ab}$ and $P^{ab}$ the ADM-momentum, given by:
\begin{align}
 P_{ab}=-\frac{1}{16 \pi G}\sqrt{q}\left( K_{ab}-q^{ab} K \right),
\end{align}
where $K_{ab}$ is the extrinsic curvature
\begin{align}
 K_{ab}=\frac{1}{2N}\left( -\dot{q}_{ab}+\nabla_b N_a + \nabla_a N_b \right).
\end{align}
Taking variations of that term of the action with respect to
the canonical variables yields the following boundary contribution, 
\begin{align}
 2 \int_{\Sigma} \mathrm{d}^3 x \nabla_b \left( N_a \delta P^{ab} \right),
 \end{align}
 which must be canceled at infinity. We must thus add to the action the following surface term:
 \begin{align}\label{eq:P-bound}
 {\cal P}=- 2 \oint_{\delta \Sigma} \mathrm{d} S^b N^a P_{ab}.
\end{align}
As we see, this boundary term not only depends on the phase space variables, but also on the Lagrange multipliers (the shift functions $N^a$ in this case). In the following, we will assume that the latter will be prescribed functions at spatial infinity (determined by the asymptotic form of the Kerr metric given below in Eq. \eqref{eq:kerr-aymp}). Therefore, we will not consider variations of these functions on the boundary. 

The boundary term in Eq. \eqref{eq:P-bound} can be easily evaluated for the Kerr metric at spatial infinity. In spherical coordinates, its asymptotic form is given by
\begin{equation}\label{eq:kerr-aymp}
 \mathrm{d}S^2 = -\left[ 1-\frac{2m}{r} \right] \mathrm{d}t^2 - \frac{2J \sin^2 \theta}{r}\left(\mathrm{d}t \mathrm{d}\phi + \mathrm{d}\phi \mathrm{d}t \right) + \left[1 + \frac{2m}{r} \right] \mathrm{d}r^2  + r^2 (\mathrm{d}\theta^2+ \sin^2 \theta \mathrm{d}\phi^2).
\end{equation}
Moreover,  
\begin{align}
 \mathrm{d}S^b=\mathrm{d}S n^b &= \mathrm{d}S \delta_r^b,
 %N^a &= N^\phi \delta^a_\phi
\end{align}
and $\mathrm{d}S=r^2\sin\theta {\rm d}\theta {\rm d}\phi$. From the metric in Eq. \eqref{eq:kerr-aymp}, the only contributions to the integral are
\begin{align}
{\cal P} =- 2 \oint_{\delta\Sigma} \mathrm{d}S N^\phi P_{\phi r}=-2 \oint_{\delta\Sigma} \mathrm{d}\theta \mathrm{d}\phi r^2 \sin\theta N^\phi \left( \frac{\sqrt{q}}{16\pi G} (K_{\phi r} - q_{\phi r} K ) \right),
\end{align}
where $q_{ab}$ is the spatial metric and $K^{ab}$ the extrinsic
curvature. At the boundary $\delta \Sigma$, we have
\begin{align*}
 q_{\phi r}&=0, \\
 \sqrt{q}&=r^{3/2} \sqrt{r+r_s} \sin\theta, \\
 N_\phi &= -\frac{2J \sin^2\theta}{r} ~~ \rightarrow ~~ N^\phi=-\frac{2J}{r^3}, \\
 K_{\phi r}&=\frac{3 J \sin^2 \theta}{\sqrt{r^3 (r-r_s)+4J^2 \sin^2 \theta}}, \\
 K&=0.
\end{align*}
Performing the integral, the leading term in the expansion in $1/r$ is
\begin{align}
{\cal P} = \lim_{r\to\infty}\frac{9\pi}{64 G} \frac{J^2}{r}=0.\label{6.9}
\end{align}

Now, in Ashtekar-Barbero variables, the diffeomorphism constraint of the full theory takes the form
\begin{align}
\frac{1}{16\pi G}D(\vec N)&=\frac{1}{8\pi G\beta}\int_\Sigma N^a\left(E^b_i\partial_aA^i_b-\partial_b(E^b_iA^i_a)
\right).
\end{align}
Its boundary term (for asymptotically flat spacetimes) takes the form
\begin{align}
{\cal P}= \frac{1}{8\pi G\beta} \oint_{\delta\Sigma} N^a E^b_i ~ A_a^i ~ \mathrm{d}S_b.
\end{align}
On the other hand, in our reduced theory, the reduced diffeomorphism
constraint is given by
\begin{align}
\frac{1}{8\pi G}d(\vec N)&=\frac{1}{4\pi G\beta}\int_\sigma N^a\left(\e^b_i\partial_a\a^i_b-\partial_b(\e^b_i\a^i_a)+\delta_a^\phi\delta^i_3\varepsilon_{ijk}\a^j_b\e^b_k
\right),
\end{align}
where $\sigma$ are the $r,\theta$ 2D spatial sections of our reduced theory, with topology $\mathbb R^2$. The corresponding boundary term is
\begin{align}
{\mathfrak p}= \frac{1}{4G\beta} \oint_{\delta\sigma} N^a \e^b_i ~ \a_a^i ~ \mathrm{d}s_b,
\end{align}
where
\begin{align}
 \mathrm{d}s^b=\mathrm{d}s n^b &= \mathrm{d}s \delta_r^b, \\
 %N^a &= N^\phi \delta^a_\phi
\end{align}
with ${\rm d}s=r^2\sin\theta {\rm d}\theta$. Since the only non-vanishing component of the shift is $N^\phi$, the required boundary term is
\begin{align}
{\mathfrak p}= \frac{1}{4 G\beta} \oint_{\delta\sigma} N^\phi ~ \e^r_3 \a^3_\phi ~ q_{rr} ~ r^2 \sin\theta ~ \mathrm{d}\theta .
\end{align}
Where we have chosen the triad as in \eqref{eq:triad}.
Again, we have at infinity:
\begin{align*}
 N^\phi&=-\frac{2J}{r^3}, \\
 e^r_3 &= r^2 \sin\theta, \\
 q_{rr} &= \frac{1}{1+\frac{r_s}{r}}, \\
 a_\phi^3 &=  \cos\theta + \gamma \frac{3J r \sin^2 \theta}{\sqrt{r(r+r_s)(r^3(r-r_s)+ 4 J^2 \sin^2\theta)}}.
\end{align*}
The term proportional to $\cos\theta$ integrates to zero (since it
contains the integral of $\cos\theta \sin\theta$ between $0$ and
$\pi$) while the other term is easily seen to yield the same result as
in (\ref{6.9}).

\subsection{Hamiltonian constraint}

Now we turn to the portion of the action involving the Hamiltonian constraint. In ADM variables:
\begin{align}
 \frac{1}{16 \pi G}C[N]=\frac{1}{16 \pi G}\int_\Sigma N \left( q^{-1/2} \left( P_{ab} P^{ab} - \frac{1}{2} P \right)- q^{1/2} R \right).
\end{align}
Where $R$ is the Ricci scalar. In order for the variations with respect to the dynamical variables to be well defined, it is necessary to add to the action the surface term (see e.g. \cite{thiemann_boundary}):
\begin{align}
 \label{boundary_adm_0}
{\cal E}=\frac{1}{16 \pi G} 2 \oint_{\delta \Sigma}\mathrm{d}S_d N \sqrt{q} q^{ac} q^{bd} \bar{\nabla}_{\left[c\right.} q_{b\left. \right]a}.
\end{align}
Where $\bar{\nabla}$ is the covariant derivative compatible with the
order zero of expansion in $1/r$ of the spatial metric at
infinity. This term corresponds to time translations at infinity. The
surface term actually has another contribution coming from the fact
that the Ricci tensor has second derivatives, requiring two
integration by parts. That contribution corresponds to boosts at
infinity, but due to our choice of adapted coordinates we do not allow
such boosts. 

We will evaluate the previous boundary term at spatial infinity in
order to show that it is finite. For convenience, we will introduce an
asymptotically Cartesian coordinate system with coordinates
$\left\lbrace x^a \right\rbrace$. We then expand our metric
asymptotically as $g_{ab}=\eta_{ab}+h_{ab}$, with $\eta_{ab}$ the flat
space metric and $h_{ab}$ a small perturbation around $\eta_{ab}$. We
also expand the lapse as $N=1+{\cal O}(1/r)$. Then, in the limit
$r\to\infty$, the leading contribution to the boundary term takes the
form
\begin{align}
\label{boundary_adm}
{\cal E} = \frac{1}{16\pi G} \oint_{\delta \Sigma} \left( \frac{\partial h_a^b}{\partial x^b} - \frac{\partial h_b^b}{\partial x^a} \right) \mathrm{d}S^a=\frac{r_s}{2G}.
\end{align}

%\begin{align*}
% \mathrm{d}S^2 = \left[\frac{r^2 + J^2 (\cos(\theta))^2}{J^2 + r^2 - r r_s}\right] \mathrm{d}r^2  + \left(r^2 + J^2 \cos(\theta)^2\right) \mathrm{d}\theta^2+ \left[(\sin(\theta))^2 \left(J^2 + r^2 + \frac{J^2 r r_s (\sin(\theta))^2}{r^2 + J^2 (\cos(\theta))^2}\right) \right] \mathrm{d}\phi^2 
%\end{align*}
%Performing an expansion in terms of $r$, we see that the metric components take the following form:
%\begin{align*}
% g_{rr}&=1 + \frac{r_s}{r} +  \frac{-J^2 +  r_s^2 + J^2 (\cos(\theta))^2}{r^2}+ O(1/r^3) \\
% g_{\theta\theta}&=r^2 + J^2 \cos^2 \theta \\
% g_{\phi\phi}&=r^2 \sin(\theta)^2 + J^2 \sin(\theta)^2 + \frac{J^2 r_s (\sin(\theta))^4}{r} + O(1/r^3)
%\end{align*}
%From where we can obtain the components of $\gamma$ in spherical coordinates: 
%\begin{align*}
% \gamma_{rr}&=\frac{r_s}{r} +  \frac{-J^2 +  r_s^2 + J^2 (\cos(\theta))^2}{r^2}+O(1/r^3) \\
% \gamma_{\theta\theta}&=J^2 \cos^2 \theta \\
% \gamma_{\phi\phi}&=J^2 \sin(\theta)^2 + \frac{J^2 r_s \sin(\theta)^4}{r} + O(1/r^3)
%\end{align*}
%By performing a change of variable and expressing $\gamma$ in cartesian coordinates, \eqref{boundary_adm} may be readily evaluated. After a lengthy but straightforward calculation one arrives at the result
%\begin{align*}
% \frac{r_s}{2G}
%\end{align*}
%Which is none other than the Schwarzschild mass. 

Now, a direct calculation shows that, in Ashtekar variables, the
equivalent boundary term to \eqref{boundary_adm_0} takes the following form:
\begin{align}
{\cal E}= -\frac{1}{8 \pi G\beta}\oint_{\delta \Sigma} \mathrm{d}S_a \frac{N}{\sqrt{E}} \left(E^a_i \bar{D}_b E^b_i + E^b_i \bar{D}_b E^a_i \right),
\end{align}
where, similar to the derivative $\bar{\nabla}$ defined earlier, $\bar{D}$ is the covariant derivative compatible with the order zero component of the triad at infinity. 

In our reduced theory, the boundary term is given by
\begin{align}
{\mathfrak e}= -\frac{1}{4 G\beta}\oint_{\delta \sigma} \mathrm{d}s_a \frac{N}{\sqrt{e}} \left(\e^a_i \bar{D}_b \e^b_i + \e^b_i \bar{D}_b \e^a_i \right).
\end{align}
Its evaluation in Cartesian coordinates agrees with the result given in \eqref{boundary_adm}.

\subsection{Gauss constraint}

The contribution to the action of the Gauss constraint in the full theory is given by
\begin{align}
 \frac{1}{16\pi G}G(\vec\lambda)&=\frac{1}{8\pi G\beta}\int_\Sigma \lambda^i\left(\partial_aE^a_i+\varepsilon_{ijk}A^j_aE^a_k\right).
\end{align}
The variation of this contribution also requires another boundary term in order to make the full variational problem well defined. It is given by
\begin{equation}\label{eq:Q}
{\cal Q}=\frac{1}{8\pi G\beta} \oint_{\delta \Sigma} dS_a (E^a_i-\bar{E}^a_i) \lambda^i,
\end{equation}
where $\bar{E}^a_i$ is the densitized triad at spatial
infinity. Without this term, inserting the asymptotic form of the
triad, the result is divergent. Since at spatial infinity the metric
is flat, $\bar{E}^a_i$ is independent of $M$ and $J$. Therefore, any
variational derivative of this term will be zero. Similar boundary
terms have been suggested in previous treatments
\cite{thiemann_boundary,campiglia_asymptotic}.

In our symmetry reduced theory, the reduced Gauss constraint is given by:
\begin{align}
 \frac{1}{8G}g(\vec\lambda)&=\frac{1}{4G\beta}\int_\sigma \lambda^i\left(\partial_a\e^a_i+\varepsilon_{ijk}\a^j_a\e^a_k+\varepsilon_{ijk}\delta^j_3\delta^\phi_a \e^a_k\right).
\end{align}
The boundary term takes the same form in terms of the reduced densitized triad, namely,
\begin{equation}
{\mathfrak q}=\frac{1}{4G\beta} \oint_{\delta \sigma} dS_a (\e^a_i-\bar{\e}^a_i) \lambda^i,
\end{equation}
where $\bar{\e}^a_i$ is the reduced triad at spatial infinity. 
%Integrating by parts, we are left with:
%\begin{align*}
% g(\vec\lambda)&=\frac{1}{8\pi\beta}\int_\Sigma \left( (-\partial_a \lambda^i - \epsilon_{kji} \a^i_a \lambda^i ) \e^a_i \right) + \frac{1}{8$\pi G} \oint_{\delta \Sigma} dS_a \e^a_i \lambda^i
%\end{align*}
%While $g(\vec{\lambda})$ is well defined, the two terms on the right hand side are not necesarily well defined by themselves \ref{campiglia_asymptotic}. In effect, when calculating the boundary term explicitly, we find a divergent term (which cancels the divergence coming from the volume integral) and a finite contribution, the $SU(2)$ charge:
After evaluation at spatial infinity, one gets
\begin{align}
{\mathfrak q} = \frac{5\pi^2}{64 G}  \frac{J^2}{r_{S}} .
\end{align}
We  note that it only depends on $M$ and $J$. Therefore, no new
observables appear. This is due to our choice of diagonal triads,
which ties spatial rotations generated by $J$ to internal rotations.

\section{Conclusions}
We have developed the Ashtekar--Barbero framework for axisymmetric
spacetimes. We found triads and connections adapted to the symmetry
and wrote the Gauss law, vector and Hamiltonian constraints. We showed
that the Kerr solution indeed solves the constraints and the evolution
equations. We also discussed the boundary terms needed to make the
action differentiable in a canonical treatment. This lays out a
framework to attempt a loop quantization of axially symmetric
spacetimes. These represent the most complex midisuperspaces
considered up to date. The strategy we intend to follow for
quantization is similar to the one we pursued in spherical symmetry
(\cite{Gambini:2013hna, Gambini:2014qta}). We will build spin network
states based on the reduced connection. The third component of the
connection will be represented by a point holonomy and the first two
with genuine holonomies in the two dimensional reduced space adapted
to the symmetry (for instance $r,\theta$ if one were to consider
spherical coordinates). On such states the fluxes of the triads will
act naturally. We will use these basic operators to construct the
Hamiltonian of the theory. We will discuss details in a future publication.

\section*{Acknowledgments}
We wish to thank Miguel Campiglia for discussions.  This work was
supported in part by grant No. NSF-PHY-1603630, funds of the Hearne
Institute for Theoretical Physics, CCT-LSU, Pedeciba, grant ANII
FCE-1-2014-1-103974, grant No.  NSF-PHY-1505411, the Eberly research funds
of Penn State, Project. No. MINECO FIS2014-54800-C2-2-P from Spain and
its continuation Project. No. MINECO FIS2017-86497-C2-2-P.

\appendix

\section{Axisymmetric classical reduction}\label{app:red}

From the components of Eq. \eqref{eq:axi-sym} we get the set of differential equations
\begin{equation}
\partial_\phi A^1_a = -A^2_a, \quad \partial_\phi A^2_a = A^1_a, \quad \partial_\phi A^3_a = 0.
\end{equation}
From $\partial_\phi A^3_a = 0$ we conclude that $A^3_a$ must be equal to a function $\a^3_a$ independent of $\phi$. On the other hand, the most general solutions for the differential equations of $A^1_a$ and $A^2_a$ are
\begin{equation}
A^1_a = \cos\phi \,\a^1_a-\sin\phi\, \a^2_a,\quad A^2_a=\sin\phi\, \a^1_a+\cos\phi\, \a^2_a,
\end{equation}
with $\a^1_a$ and $\a^2_a$ independent of $\phi$. These solutions yield to Eq. \eqref{eq:red-A}. The symmetry reduction of the densitized triad $E^a_i$ (despite being a tensor density of weight one) is similar to the one of $A^i_a$ for axisymmetric spacetimes (for $\lambda_3={\rm const}$). The reduced components of $E^a_i$ take a similar form as those of $A^i_a$ but replacing $\a^i_a$ by $\e^i_a$. One then obtains Eq. \eqref{eq:red-E}.  

\section{Kerr solution: Lagrange multipliers and connexion components}

Gauss' law Lagrange multipliers:
\begin{align*}
\lambda^1&=0, 
\\
\lambda^2&=\frac{\sqrt{2} a \sqrt{a^2+r(r-r_s)} r_s ( a^4 - 3 a^2 r^2 - 6 r^4 + 4 a r (a^2+r^2) \beta \cos(\theta)+a^2(a^2-r^2)\cos(2\theta) ) \sin(\theta) %  
}{ ( ( a^2+2 r^2 + a^2 \cos(2 \theta) ) ( a^4 + 2 r^4 + a^2 r (3r + r_s)+a^2 (a^2+r(r-r_s) ) \cos(2\theta))^{3/2} )},
\end{align*}
%\begin{align*}
%\Lambda^3&=\frac{1}{2} \gamma \sqrt{\frac{J^2+r(r-r_s)}{(J^2+r^2)(r^2+J^2 \cos(\theta)^2)+J^2 r r_s \sin(\theta)^2 }} \left( \frac{(2r-r_s)\sin(\theta)}{\sqrt{J^2+r(r-r_s)}} +  \right.
%\\
%&\frac{\left(8 J^3 r \sqrt{J^2+r(r-r_s)} r_s (1+\gamma^2)\cos\theta \sin\theta^3 \right)}{\left(\gamma (J^2+2 r^2 + J^2 \cos2\theta) (J^4+2 r^4 + J^2 r (3r+r_s)+J^2 (J^2 + r (r-r_s)) \cos2\theta ) \right)}
%\end{align*}
\begin{align*}
&\lambda^3=\left(\frac{a^2 + r (r - r_s)}{(a^2 + r^2) (r^2 + a^2 \cos^2(\theta)) +
a^2 r r_s \sin^2(\theta)}\right)^{1/2} \left\lbrace\frac{(2 r - r_s) \sin(\theta)}{(a^2 + r (r - r_s))^{1/2}} \right.
\\
&+ \frac{8 a^3 r (a^2 + r (r -
r_s))^{1/2} r_s (1 + \beta^2) \cos(\theta) \sin^3(\theta)}{\beta \
(a^2 + 2 r^2 + a^2 \cos(2 \theta)) (a^4 + 2 r^4 + a^2 r (3 r + r_s) + \
a^2 (a^2 + r (r - r_s)) \cos(2 \theta))} 
\\
&- \left[ \frac{4 (r^2 + a^2
\cos^2(\theta)) ((a^2 + r^2) (r^2 + a^2 \cos^2(\theta)) + a^2 r
r_s (\sin(\theta))^2)^{1/2}}{{a^4 + 2 r^4 + a^2 r (3 r + 
r_s) + a^2 (a^2 + r (r - r_s)) \cos(2 \theta)}} \right.
\\
& \left( \frac{(16 r^5 + 4 a^2 r^2 (4 r - r_s) +
a^4 (6 r + r_s) + 4 a^2 r (2 a^2 + r (4 r + r_s)) \cos(2 \theta) + a^4
(2 r - r_s) \cos(4 \theta)) \sin(\theta)}{8 \sqrt{2} (r^2 + a^2
\cos^2(\theta))^2 \left(\frac{a^4 + 2 r^4 + a^2 r (3 r + r_s)}{a^2 + r (r
- r_s)} + a^2 \cos(2 \theta)\right)^{1/2}} \right.
\\
&\left. \left. \left.
+\frac{8 a^3 r r_s \beta \cos(\theta) \sin^3(\theta) ((a^2 + r (r - r_s)) ((a^2 + r^2) (r^2 +
a^2 \cos^2(\theta)) + a^2 r r_s \sin^2(\theta)))^{1/2}}{(a^2 + 2
r^2 + a^2 \cos(2 \theta))^2 (a^4 + 2 r^4 + a^2 r (3 r + r_s) + a^2
(a^2 + r (r - r_s)) \cos(2 \theta))}\right)\right]\right\rbrace.
\end{align*}

Connection components:
\begin{align*}
 &a_r^1=\frac{-2 a r_s \beta (r^2 + a^2 \cos^2(\theta)) (a^4 -3 a^2 r^2 -6 r^4 + a^2 (a - r) (a + r) \cos(2 \theta)) \sin(\theta)}{(a^2 + r (r - r_s))^{1/2} (a^2 + 2 r^2 + a^2 \cos(2 \theta))^2 (a^4 + 2 r^4 + a^2 r (3 r + r_s) + a^2 (a^2 + r (r - r_s)) \cos(2 \theta))}
 \\
 &+ \frac{a^2 \sin(2 \theta)}{(a^2 + r (r - r_s))^{1/2} (a^2 + 2 r^2 + a^2 \cos(2 \theta))}, \\
 &a_r^2=0, \\
 &a_r^3=0, \\
 &a_\theta^1=\frac{r (a^2 + r (r - r_s))^{1/2}}{r^2 + a^2 \cos^2(\theta)} 
 \\
 &- \frac{8 a^3 r (a^2 + r (r - r_s))^{1/2} r_s \beta \cos(\theta) (r^2 + a^2 \cos^2(\theta)) \sin^2(\theta)}{(a^2 + 2 r^2 + a^2 \cos(2 \theta))^2 (a^4 + 2 r^4 + a^2 r (3 r + r_s) + a^2 (a^2 + r (r - r_s)) \cos(2 \theta))}, \\
 &a_\theta^2=0, \\
 &a_\theta^3=0, \\
 &a_\phi^1= 0, \\
 &a_\phi^2\!=\! \frac{(16 r^5 + 4 a^2 r^2 (4 r - r_s) + a^4 (6 r + r_s) + 4 a^2 r (2 a^2 + r (4 r + r_s)) \cos(2 \theta) + a^4 (2 r - r_s) \cos(4 \theta)) \sin(\theta)}{8 \sqrt{2} (r^2 + a^2 \cos^2(\theta))^2 (\frac{a^4 + 2 r^4 + a^2 r (3 r + r_s)}{a^2 + r (r - r_s)} + a^2 \cos(2 \theta))^{1/2}}
 \\
 &+ \frac{8 a^3 r r_s \beta \cos(\theta) \sin^3(\theta) ((a^2 + r (r - r_s)) ((a^2 + r^2) (r^2 + a^2 \cos^2(\theta)) + a^2 r r_s \sin^2(\theta)))^{1/2}}{(a^2 + 2 r^2 + a^2 \cos(2 \theta))^2 (a^4 + 2 r^4 + a^2 r (3 r + r_s) + a^2 (a^2 + r (r - r_s)) \cos(2 \theta))},
\end{align*}
\begin{align*}
 &a_\phi^3=\frac{2 (a^2 (a^2 + r (r - r_s)) ((5 a^2 + 8 r^2)
   \cos(3 \theta) + a^2 \cos(5 \theta))}{2 \sqrt{2} (a^2 + 2 r^2 + a^2
   \cos(2 \theta))^2 (a^4 + 2 r^4 + a^2 r (3 r + r_s) + a^2 (a^2 + r (r
   - r_s)) \cos(2 \theta))^{1/2}} \\
&+\frac{2 (5 a^6 + 8 r^6 + 4 a^2 r^3 (5 r + r_s) + a^4 r (17 r + 3 r_s)) \cos(\theta) }{2 \sqrt{2} (a^2 + 2 r^2 + a^2 \cos(2 \theta))^2 (a^4 + 2 r^4 + a^2 r (3 r + r_s) + a^2 (a^2 + r (r - r_s)) \cos(2 \theta))^{1/2}} \\
 &+\frac{\sqrt{2} a r_s \beta (- a^4 + 3 a^2 r^2 + 6 r^4 + a^2 (- a^2 + r^2) \cos(2 \theta)) \sin^2(\theta)}{(a^2 + 2 r^2 + a^2 \cos(2 \theta))^2 (a^4 + 2 r^4 + a^2 r (3 r + r_s) + a^2 (a^2 + r (r - r_s)) \cos(2 \theta))^{1/2}}.
\end{align*}

\end{document}